\documentclass{aims}
\usepackage{amsmath}
  \usepackage{paralist}
  \usepackage{graphics} 
  \usepackage{epsfig} 
\usepackage{graphicx}  \usepackage{epstopdf}
 \usepackage[colorlinks=true]{hyperref}
\hypersetup{urlcolor=blue, citecolor=red}

\def\currentvolume{X}
 \def\currentissue{X}
  \def\currentyear{200X}
   \def\currentmonth{XX}
    \def\ppages{X--XX}
     \def\DOI{10.3934/xx.xx.xx.xx}

\theoremstyle{definition}

\title[Travelling wave solutions of  glioblastoma growth] 
      {Travelling wave solutions of the reaction-diffusion mathematical
model of glioblastoma growth: An Abel equation based approach}

\author[T. Harko and M. K.  Mak]{}


 \email{t.harko@ucl.ac.uk}
 \email{mkmak@vtc.edu.hk}

\begin{document}
\maketitle

\centerline{\scshape Tiberiu Harko }
\medskip
{\footnotesize
 \centerline{Department of Mathematics, University College London, Gower Street, London
WC1E 6BT, United Kingdom}

} 

\medskip

\centerline{\scshape M. K.  Mak }
\medskip
{\footnotesize
 \centerline{ Department of Computing and Information Management, Hong Kong Institute of
Vocational Education,}
\centerline{ Chai Wan, Hong Kong, P. R. China}
}

\bigskip


\begin{abstract}
We consider quasi-stationary (travelling wave type) solutions to a nonlinear
reaction-diffusion equation with arbitrary, autonomous coefficients,
describing the evolution of glioblastomas, aggressive primary brain tumors
that are characterized by extensive infiltration into the brain and are
highly resistant to treatment. The second order nonlinear equation
describing the glioblastoma growth through travelling waves can be reduced to a first order Abel type
equation. By using the integrability conditions for the Abel equation
several classes of exact travelling wave solutions of the general reaction-diffusion equation that describes glioblastoma growth are obtained, corresponding to different forms of the product of the diffusion and reaction functions.  The solutions are obtained by using the Chiellini lemma and the Lemke transformation, respectively, and the corresponding equations represent  generalizations of the classical Fisher--Kolmogorov equation. The biological implications of two classes of solutions are also investigated by using both numerical and semi-analytical methods for realistic values of the biological parameters.
\end{abstract}

\section{Introduction}

 Glioblastoma, the most common central nervous system (CNS) tumor, accounting for 50\% of the 17,000 primary brain tumors
diagnosed annually in the US \cite{bp}, is also the most malignant form of brain cancer, having an extremely poor outcome \cite{bp1}. For
the most aggressive grade of gliomas, known as glioblastoma multiforme (GBM),
the life expectancies are from 6 to 12 months \cite{1a}. Amongst patients treated with
surgery and a radiation-containing regimen, median survival
was 12.0 months in the period 2000--2003, and 14.2 months in
2005--2008, respectively. In the temozolomide era, median survival
times varied from a high of 31.9 months, for patients in the age group
20--29 years, to a low of 5.6 months in patients age 80 years, and older \cite{sur}.

One factor that
makes glioblastoma extremely difficult to treat is its high invasiveness,
enabling tumor cells to disperse from the main tumor mass into the
surrounding normal brain. Glioblastoma is highly diffuse, and can invade a
large portion of the cerebral cortex in a short period of time, making
complete surgical excision impossible, so that dispersed glioma cells are
out of reach of surgery, radiation, and chemotherapy, so that recurrence
becomes inevitable \cite{2}. There are many factors determining the
prognosis for patients with gliomas, like the histologic type, the grade of
malignancy, the patient's age and the level of neurological functioning,
respectively \cite{2a}. The grade of malignancy for glioblastoma includes at
least two factors, the net proliferation rate, and the invasiveness,
respectively, which are estimated histologically  (for the World Health Organization (WHO)  classification and grading of brain tumors see \cite{WHO}). However, a practical accurate
definition of these factors is still missing \cite{3}. Unlike solid tumors,
for which simple exponential or geometric expansion represents expansion of
volume (equivalent to the number of cells in the tumor), gliomas consist of
motile cells that can migrate as well as proliferate \cite{3}. Indeed, the
invasiveness makes it almost impossible to define the growth rate as a
classical volume-doubling time (DT) \cite{3}.  DT is a parameter
widely used for the measurement of the tumor growth rate, which can also be
quantified by the specific growth rate (SGR), defined as follows. If the tumor volume $V$ is measured at times
$t_1$ and $t_2$, then SGR can be obtained as ${\rm SGR}=\ln\left(V_2/V_1\right)/\left(t_2-t_1\right)$ \cite{DT}. DT is related to SGR by the relation ${\rm DT}=\ln2/{\rm SGR}$ \cite{DT}.

Cancer research has been a fertile ground for developing mathematical models
that can describe the proliferation of malignant cells. The early models of
cell proliferation were based on a simple exponential growth of solid
(usually benign) tumors, so that
\begin{equation}
n(t)=N_0\exp (\lambda t),
\end{equation}
where $n$ is
the number of cells at time $t$, $N_0$ is the initial cell number, and $\lambda $ is a constant \cite{E}. Another model used to
describe tumor dynamics is based on the Gompertz curve, a mathematical model
for a time series, where growth is slowest at the end of a time period,
\begin{equation}
n(t)=N_0\exp\left\{\ln\left[\frac{N_{\infty}}{N_0}\left(1-\exp (-bt)\right)%
\right]\right\},
\end{equation}
where $N_{\infty}$ is the plateau cell number, which is reached at large
values of the time. The parameter $b$ is related to the initial tumor
growth rate \cite{4}. An alternative model describing the time variation of
a mass $m$ of any organism, including solid tumors, is given by \cite{5}
\begin{equation}
\frac{dm}{dt}=am^{3/4}-bm,
\end{equation}
where $a,b=\mathrm{constant}$. However, such models do not take into account
the spatial arrangement and distribution of the cells at a specific
anatomical location, or the spatial spread of the cancerous cells. These
spatial aspects are crucial in estimating tumor growth, since they
determine the invasiveness of the tumor and the sharpness of the apparent
border of the tumor \cite{6}. Explicitly taking into account the extent of
infiltration of the tumor is necessary in different situations, like, for
example, in estimating the benefit of surgical resection.

A simple mathematical model for the proliferation and infiltration of the
glioma cells was introduced in \cite{6}. The model was based in part on
quantitative image analysis of histological sections of a human brain glioma
and especially on cross-sectional area/volume measurements of serial
Computer Tomograph (CT) images while the patient was undergoing
chemotherapy. In its general form, from a mathematical point of view the
model represents a reaction-diffusion system, with the growth rate and the
diffusion rate representing the key model parameters. An extension of the
mathematical model based on proliferation and infiltration of neoplastic
cells introduced in \cite{6} that allows predictions to be made concerning
the life expectancies following various extents of surgical resection of
gliomas of all grades of malignancy was considered in \cite{7}. Numerical
simulations using the model allow to estimate what would happen to patients
if various extents of surgical resection, rather than chemotherapies, would
have been used. It has been shown that the shell of the infiltrating tumor
that remains after 'gross total removal' or even a maximal excision
continues to grow and regenerates the tumor mass remarkably rapidly \cite{7}%
. The model initially introduced in \cite{5,6} has been extended and used in
different clinical situations for the mathematical study of glioma
proliferation and invasion (for very informative reviews see \cite{3} and
\cite{8}) and to include the effects of radiotherapy \cite{9}, where an
extension to Swanson's reaction-diffusion model to include the effects of
radiation therapy using the classic linear-quadratic radiobiological model
was presented. Moreover, in \cite{8} it was shown that the defining and
essential characteristics of gliomas in terms of net rates of proliferation
and invasion can be determined from serial Magnetic Resonance Images (MRIs)
of individual patients. To gain some insight into glioblastoma invasion,
experiments were conducted on the patterns of growth and dispersion of U87 glioblastoma tumor spheroids in a three-dimensional collagen gel in \cite{1}.
 A continuum mathematical model of the dispersion
behaviors was developed. The mathematical
model quantitatively reproduces the experimental data.

Chemotherapy in a spatial model of tumor growth was
considered in \cite{9a}. The model, which is of reaction-diffusion type,
takes into account the complex interactions between the tumor and
surrounding stromal cells by including densities of endothelial cells and
the extra-cellular matrix. When no treatment is applied the model reproduces
the typical dynamics of early tumor growth. A mathematical model using a
realistic three-dimensional brain geometry, and which and considers
migrating and proliferating cells as separate classes was analyzed in \cite%
{9b}. Several mechanisms for infiltrative migration were considered, and
methods for simulating surgical resection, radiotherapy and chemotherapy
were developed. It was shown that the model provides clinically realistic
predictions of tumor growth and recurrence following therapeutic
intervention. An important aspect of the biological modeling of glioblastoma is the mathematical handling of boundary conditions.
An explicit and thorough numerical formulation of the adiabatic
Neumann boundary conditions imposed by the skull on the diffusive growth of gliomas
and in particular on glioblastoma multiforme was considered in \cite{9b1}. A detailed
exposition of the numerical solution process for a homogeneous approximation of glioma
invasion using the Crank--Nicolson technique in conjunction with the Conjugate Gradient
system solver was also provided.

 An individual-based stochastic model that analyses how the phenotypic switching between proliferative and migratory states of individual cells affects the macroscopic growth of the tumour was proposed in \cite{Gerlee}. The glioblastoma cells are either in a proliferative state, where they are stationary and divide, or in motile state, in which they are subject to random motion. This model may find some clinical applications for designing relevant cell screens for glioblastoma and cytometry-based patient prognostic.  

The simple reaction-diffusion model of gliomas \cite{6,7} predicts that the
"edge" of the detectable glioma should behave as a travelling wave and
follow Fisher's approximation that the velocity equals twice the square root
of the product of the growth rate and of the diffusion coefficient. This
predictions has been confirmed by using data derived from medical imaging
techniques \cite{9c}. The implication of this result is that the prognosis
of any individual patient can be predicted if two sets of scans, separated
in time so that a significant increase in growth can be measured, can be
obtained before treatment is begun \cite{9c}.

Reaction-diffusion equations usually do admit travelling wave solutions \cite{9cm,9c0}. It
has been shown in \cite{9c1} that wave solutions, which correspond to moving
bands of concentration do exist in the system of equations describing the
Belousov-Zhabotinskii reaction. Various relevant properties of the solutions
have also been obtained. With parameter values obtained from experiment,
numerical results have been given for the travelling wave solutions. A semi-inverse method, which renders exact static solutions of one-component, one-dimensional reaction-diffusion  equations with variable diffusion coefficient $D$, requiring at most qualitative
information on the spatial dependence of the latter was introduced in \cite{9c01}. Through a simple ansatz the reaction-diffusion  equations can be mapped onto the stationary Schr\"{o}dinger equations, having the form of the potential still at our disposal.
A new
substitution was used in \cite{9c2} to reduce the problem of a
quasi-stationary solution to a nonlinear reaction-diffusion equation with
arbitrary, autonomous coefficients to either a linear ordinary differential
equation, or the first order first kind Abel differential equation of the form
\begin{equation}
y'=f_0(x)+f_1(x)y+f_2(x)y^2+f_3(x)y^3,
\end{equation}
where $f_i(x)$, $i=0,1,2,3$, are arbitrary real functions of $x$, defined on a real interval $I\subseteq \Re $, with $f_0,f_1,f_2,f_3\in
C^{\infty }(I)$ \cite{Golub}. The second kind Abel differential equation is defined as
\begin{equation}
\left[g_0(x)+g_1(x)y\right]y'=f_0(x)+f_1(x)y+f_2(x)y^2+f_3(x)y^3,
\end{equation}
where $g_i(x)$, $i=0,1$ are real functions of $x$. The second kind Abel differential equation can be reduced to  the first kind Abel equation by means of the substitution $g_0(x)+g_1(x)y=1/z$ \cite{Golub}.

Exact periodical and solitary wave
solutions were obtained by solving the corresponding Abel equations \cite{9c2}. In particular, it was shown that the
problem of the kinetics of thin film growth (or of wire-like nanostructures)
has bounded solutions in terms of elliptic functions. A direct method to obtain travelling-wave solutions
of some nonlinear partial differential equations expressed in terms of solutions of the
Abel differential equation of the first type with constant coefficients was proposed in \cite{9c3}.   Exact solutions to the modified Benjamin, Bona, and Mahony (BBM) equation by viscosity were found in \cite{9c4}, by including the effect of a small dissipation on waves. Using Lyapunov functions and dynamical systems theory, it was proven that when viscosity is added to the BBM equation, in certain regions there still exist bounded travelling wave solutions in the form of solitary waves, periodic, and elliptic functions.

Travelling-wave solutions of different mathematical model describing the
growth of tumors have been considered in several publications. Spreading
cell fronts play an essential role in many physiological and biological
processes. Classically, models of this process are based on the Fisher -
Kolmogorov - Petrovsky - Piscounov equation (Fisher - Kolmogorov equation for short) \cite{Fish, Fish1, Kolm}; however, such continuum
representations are not always suitable as they do not explicitly represent
behaviour at the level of individual cells. Additionally, many models examine
only the large time asymptotic behaviour, where a travelling wave front with
a constant speed has been established \cite{9d}.

A travelling--wave analysis
of a mathematical model describing the growth of a solid tumor in the
presence of an immune system response was analyzed in \cite{10}. From a
modelling perspective, attention was focused upon the attack of tumor cells
by Tumor Infiltrating Cytotoxic Lymphocytes (TICLs), in a small
multicellular tumor, without necrosis and at some stage prior to
(tumor-induced) angiogenesis. The existence of travelling-wave solutions
for the system was established.

 A lattice-gas cellular automaton model of tumor cell proliferation, necrosis and tumor cell
migration was introduced in \cite{Hatz}, with the main aim of predicting the velocity of the traveling invasion front,
which depends upon fluctuations that arise from the motion of the discrete cells at the
front. An analytical estimate of the velocity was derived in the cut-off mean-field approximation via
the discrete Lattice Boltzmann equation and its linearization. The front velocity
scales with the square root of the product of probabilities for mitosis and the
migration coefficient, while the width of the traveling front is found
to be proportional to its velocity. In \cite{Hatz1} it was shown, with the help of a simple growth model,  that the short time required for the
recurrence of a glioblastoma multiforme tumour after a gross total resection cannot be explained solely by a mutation-based
theory. It was proposed that the transition to invasive tumour phenotypes can be explained on the basis of the
microscopic ‘Go or Grow’ mechanism (migration/proliferation dichotomy) and the oxygen shortage, i.e.
hypoxia, in the environment of a growing tumour. This hypothesis was tested with  the help of the lattice-gas
cellular automaton model \cite{Hatz}. Possible therapies that could help prevent the progression towards
malignancy and invasiveness of benign tumours were also suggested. A mathematical model that incorporates the interplay among two tumor cell phenotypes, a
necrotic core and the oxygen distribution, and which reveals the formation of
a traveling wave of tumor cells was analyzed in \cite{Alicia}. The model reproduces the observed histologic patterns of pseudopalisades.  The simulations of the model equations also show that preventing the collapse of tumor microvessels leads to slower glioma invasion.

A cell-based model of glioblastoma growth,
which is based on the assumption that the cancer cells switch phenotypes
between a proliferative and motile state, was analyzed in \cite{11}. The
dynamics of this model can be described by a system of partial differential
equations, which exhibits travelling wave solutions whose wave speed depends
crucially on the rates of phenotypic switching. Under certain conditions on
the model parameters, a closed form expression of the wave speed can be
obtained. By using singular perturbation methods an approximate expression
of the wave front shape can be derived.

A simple free boundary model formed
of a Hele-Shaw equation for the cell number density coupled to a diffusion
equation for a nutrient was studied in \cite{12}. In this model a travelling
wave solution does exist, with a healthy region separated from the
progressing tumor by a sharp front (the free boundary), while the transition
to the necrotic core is smoother. The pressure distribution vanishes at the
boundary of the proliferative rim, with a vanishing derivative at the
transition point to the necrotic core.

The Fisher--Kolmogorov equation belongs to a more general class of reaction-diffusion equations, given by \cite{9cm,9c0}
\begin{equation}\label{gen}
\frac{\partial u}{\partial t}=\frac{\partial ^2u^m}{\partial x^2}+\frac{\partial }{\partial x}\left[\left(b_0+b_1u^p\right)u\right]+u^{2-m}\left(1-u^p\right)\left(c_0+c_1u^p\right),u>0,
\end{equation}
where $m$, $p$, $b_0$, $b_1$, $c_0$ and $c_1$ are constants. The Fisher--Kolmogorov equation is a special case of Eq.~(\ref{gen}) for $m=p=1$, $b_0=b_1=0$, $c_0=1$, and $c_1=0$, or, alternatively, $m=1$, $p=1/2$, $b_0=b-1=0$, and $c_0=c_1=1$.

It is the goal of the present paper to study exact traveling wave solutions
in the one-dimensional reaction-diffusion models of glioblastoma tumor
growth introduced in \cite{6}-\cite{9}. More exactly, we will consider the
possibility of the description of the tumor growth by a general
reaction-diffusion system, containing two arbitrary tumor concentration dependent functions, called the diffusion function  and the reaction function, respectively. By considering travelling wave solutions of the
reaction-diffusion equation, we show that the second order non-linear
differential equation describing wave propagation can be reduced to a first
kind non-linear Abel equation. Note that the mathematical properties of the Abel
equation and its applications have been intensively investigated in \cite{18}%
-\cite{HLM}.

By using some standard integrability conditions of the
Abel equation, we obtain several classes of exact solutions of the tumor growth
equation. More exactly, the first class of solutions is obtained by using the integrability condition of Chiellini \cite{Chiel, kamke}, which can be formulated as a differential condition relating the diffusion and the reaction functions. The use of this condition allows to obtain exact travelling wave solutions to some reaction-diffusion equations representing a generalization of the Fisher--Kolmogorov equation. The second class of solutions is obtained by transforming the Abel equation to a second order equation \cite{kamke, Lemke}, which allows the construction of exact travelling wave solutions for several choices of the diffusion and reaction functions. The biological implications of two tumor growth models by travelling wave propagation, both representing generalizations of the Fisher--Kolmogorov model, are investigated in detail by using some realistic numerical values for the free parameters of the model.

The present paper is organized as follows. In Section~\ref{sect2} we briefly
review the basic mathematical model of glioblastoma growth, and the
reduction of the general diffusion-reaction equation to an Abel equation is
presented. Exact solutions of the glioblastoma growth equation based on the
Chiellini lemma are presented in Section~\ref{sect3}. Further integrability
cases  of the tumor growth equation by diffusion are
obtained in Section~\ref{sect4}. A number of exact travelling wave solutions of the tumor growth equation, corresponding to different functional forms of the product of the diffusion and reaction functions are presented in Section~\ref{new}.  The biological implications of our models are briefly investigated in Section~\ref{sect4a}. We discuss and conclude our results in Section~\ref{sect5}.

\section{The mathematical model of tumor growth}

\label{sect2}

The first model of the growth of an infiltrating glioma as a
reaction-diffusion system was initially formulated as a conservation
equation \cite{13}. From a phenomenological point of view, the model can be
formulated in words as \cite{6}-\cite{9}, \cite{13}:

\begin{center}
{\it "The rate of change of tumor cell population = the diffusion
(motility) of tumor cells + the net proliferation of tumor cells}."
\end{center}

Mathematically, the growth of the gliomas can be described as a diffusion
equation for the tumor cell density $c\left(\vec{r},t\right)$ at the
position $\vec{r}$ and time $t$,
\begin{equation}  \label{1}
\frac{\partial c}{\partial t}=\nabla \cdot \vec{J}+\rho c,
\end{equation}
where $\vec{J}$ is the cell flux, and $\rho $ denotes the net proliferation
rate. By assuming that the flux obeys the standard Fick law, $\vec{J}%
=D\nabla c$, where $D$ is the diffusion coefficient representing the active
motility of glioma cells, Eq.~(\ref{1}) can be written as a
reaction-diffusion equation,
\begin{equation}
\frac{\partial c}{\partial t}=\nabla \cdot \left(D\nabla c\right)+\rho c.
\end{equation}
The model formulation is completed by boundary conditions, which impose no
migration of cells beyond the brain boundary, and initial conditions $c\left(%
\vec{r},0\right) = f \left(\vec{r}\right)$, where $f \left(\vec{r}\right)$
defines the initial spatial distribution of malignant cells. As boundary
conditions, it is required that there is no flux of cells to the outside of
the brain, or into the ventricles, so at the boundaries of the
two-dimensional domain $\vec{n}\cdot \nabla c=0$, where $\vec{n}$ is the
unit vector normal to the boundary.

However, more general models can also be formulated. In \cite{6} and \cite{7}
glioblastoma modelling was done by using the Fisher - Kolmogorov equation
\cite{Fish,Fish1,Kolm},
\begin{equation}\label{FK}
\frac{\partial c}{\partial t}=D\frac{\partial ^{2}c}{\partial x^{2}}+\rho
c\left( 1-\frac{c}{c_{max}}\right) ,
\end{equation}%
which describes randomly moving cells, and which simultaneously divide at
rate $\rho $. Cells throughout the tumor are assumed to proliferate at a
constant rate $\rho $ until they reach a limiting density $c_{max}$. The
Fisher equation exhibits travelling wave solutions, which from biological
point of view describe a tumor invading the healthy tissue, with the
velocity of the invading front given by $v=2\sqrt{D\rho }$ \cite{11}. The
inclusion of the chemotherapy in the model leads to a modification of the
free evolution of the cells according to the phenomenological prescription \cite{6}:

\begin{center}
{\it
"The rate of change of tumor cell population = diffusion (motility) of tumor cells
+ net proliferation of tumor cells - loss of tumor cells due to chemotherapy."}
\end{center}

Mathematically, the simplest model of glioblastoma evolution in the presence
of chemotherapy is formulated as
\begin{equation}\label{chem}
\frac{\partial c}{\partial t}=\nabla \cdot \left( D\nabla c\right) +\rho
c-G(t)c,
\end{equation}%
where $G(t)$ is a function describing the effects of chemotherapy. When
chemotherapy is administered, $G(t)$ is constant, and $G(t)=0$ otherwise.
The effect of the radiotherapy can be modelled as
\begin{equation}\label{rad}
\frac{\partial c}{\partial t}=\nabla \cdot \left( D\nabla c\right) +\rho
c-R\left(\vec{r},t\right)c,
\end{equation}
where $R\left(\vec{r}, t\right)$ represents the effect of external beam
radiation therapy at location $\vec{r}$ and time $t$ \cite{9}.

In the following we propose to model the evolution of gliomas by means of
1+1 dimensional general reaction - diffusion system of the form
\begin{equation}
\frac{\partial c(x,t)}{\partial t}=\frac{\partial }{\partial x}\left[
D(c(x,t)\frac{\partial c(x,t)}{\partial x}\right] +Q(c(x,t)),  \label{2}
\end{equation}%
where the dissipative (diffusion) function $D(c)\neq 0$ and the
reaction term $Q(c)\neq 0$ both depend explicitly upon the cell density
$c$ only, and not on space $x$ and time $t$ variables.

By introducing a
phase variable
\begin{equation}
\xi =x-V_{f}t,
\end{equation}
where $V_{f}\geq 0$ is a constant wave
velocity, Eq.~(\ref{2}) takes the form of a second order non-linear
differential equation of the form
\begin{equation}
\frac{d^{2}c}{d\xi ^{2}}+f(c)\left( \frac{dc}{d\xi }\right) ^{2}+g(c)\frac{dc%
}{d\xi }+h(c)=0,  \label{3}
\end{equation}%
where
\begin{equation}
f(c)=\frac{d}{dc}\ln D(c),
\end{equation}
\begin{equation}
g(c)=\frac{V_{f}}{D(c)},
\end{equation}
\begin{equation}
h(c)=\frac{Q(c)}{D(c)}.
\end{equation}
Eq.~(\ref{3}) must be considered together with the initial conditions $%
c(0)=c_0$ and $\left.\left(dc/d\xi\right)\right |_{\xi =0}=c_0^{\prime }$,
respectively. By means of the transformations
\begin{equation}
\frac{dc}{d\xi }=u,\frac{d^{2}c}{d\xi ^{2}}=u\frac{du}{dc},u=\frac{1}{v},
\end{equation}%
Eq.~(\ref{3}) can be transformed to the general form of the first order
first kind Abel equation, given by
\begin{equation}
\frac{dv}{dc}=f(c)v+g(c)v^{2}+h(c)v^{3}.  \label{4}
\end{equation}
Eq.~(\ref{4}) must be integrated with the initial condition $%
v\left(c_0\right)=1/u\left(c_0\right)=1/c_0^{\prime }$.

By introducing a new variable $w$, defined as
\begin{equation}
v=e^{\int {f(c)dc}}w=D(c)w,
\end{equation}%
allows us to write Eq.~(\ref{4}) in the standard form of the Abel equation,
\begin{equation}
\frac{dw}{dc}=V_{f}w^{2}+Q(c)D(c)w^{3},  \label{5}
\end{equation}%
which must be solved with the initial condition
\begin{equation}  \label{inw}
w\left( c_{0}\right) =w_{0}=\frac{1}{c_{0}^{\prime }D\left( c_{0}\right) }.
\end{equation}

\section{The Chiellini integrability condition for general
reaction-diffusion systems}

\label{sect3}

An exact integrability condition for the Abel equation Eq.~(\ref{5}) was
obtained by Chiellini \cite{Chiel} (see also \cite{HLM} and \cite{kamke}),
and can be formulated as the Chiellini Lemma as follows: 

\textbf{Chiellini Lemma}. \textit{If the coefficients $F(x)$ and $L(x)$ of a
first kind Abel type differential equation of the form}
\begin{equation}
\frac{dv}{dx}=F(x)v^{2}+L(x)v^{3}.
\end{equation}%
\textit{\ satisfy the condition }%
\begin{equation}
\frac{d}{dx}\left[ \frac{L(x)}{F(x)}\right] =kF(x),
\end{equation}
\textit{where $k=\mathrm{constant}\neq 0$, then the Abel Eq.~(\ref{4}) can
be exactly integrated}. 

As applied to the Abel Eq.~(\ref{5}), the Chiellini lemma states that if the
coefficients of the equation satisfy the condition
\begin{equation}
D(c)Q(c)=kV_{f}^{2}c,  \label{cond}
\end{equation}%
the Abel equation is exactly integrable. In this case the Abel Eq.~(\ref{5})
takes the form
\begin{equation}
\frac{dw}{dc}=V_{f}w^{2}+kV_{f}^{2}cw^{3}.  \label{7}
\end{equation}%
With the help of the substitution
\begin{equation}
w=\frac{1}{V_{f}}\frac{W}{c},
\end{equation}%
Eq.~(\ref{7}) becomes
\begin{equation}
c\frac{dW}{dc}=W\left( 1+W+kW^{2}\right) ,  \label{W}
\end{equation}%
which must be integrated with the initial condition
\begin{equation}
W\left( c_{0}\right) =W_{0}=\frac{V_{f}c_{0}}{c_{0}^{\prime }D\left(
c_{0}\right) }.
\end{equation}

Eq.~(\ref{W}) has the general solution
\begin{equation}
c(W,k)=C^{-1}e^{H(W,k)},
\end{equation}%
where $C^{-1}$ is an arbitrary constant of integration,
\begin{equation}
H(W,k)=\int {\frac{dW}{W\left( 1+W+kW^{2}\right) }},
\end{equation}%
and
\begin{equation}
e^{H(W,k)}=\left\{
\begin{array}{lll}
\frac{W}{\sqrt{kW^{2}+W+1}}\exp \left[ -\frac{1}{\sqrt{4k-1}}\arctan \left(\frac{%
1+2kW}{\sqrt{4k-1}}\right)\right] ,\qquad k>\frac{1}{4}, &  &  \\
&  &  \\
\frac{W}{W+2}\exp \left( \frac{2}{W+2}\right) ,\qquad k=\frac{1}{4},\label%
{sol2} &  &  \\
&  &  \\
\frac{W}{\sqrt{kW^{2}+W+1}}\exp \left[ -\frac{1}{\sqrt{1-4k}}{\rm arctanh}\left( \frac{%
1+2kW}{\sqrt{1-4k}}\right)\right] ,\qquad k<\frac{1}{4}, &  &   \\
\end{array}%
\right.
\end{equation}%
respectively. In order to obtain the explicit form of the function $H(W,k)$
we have used the algebraic identities
\begin{eqnarray}
\frac{1}{W\left( a+bW+kW^{2}\right) }&=&\frac{1}{aW}-\frac{1}{2a}\left( \frac{%
2kW+b}{a+bW+kW^{2}}\right) -\nonumber\\
&&\frac{b}{2ak}\left[ \frac{1}{\left(
W+b/2k\right) ^{2}+\left( 4ak-b^{2}\right) /4k^{2}}\right] ,  \label{ss}
\end{eqnarray}%
where $a$ and $b$ are arbitrary constants, with the particular condition $%
a=b=1$, and
\begin{equation}
\frac{4}{W(W+2)^{2}}=\frac{1}{W}-\frac{1}{W+2}-\frac{2}{(W+2)^{2}},
\end{equation}%
respectively.

Therefore Eq.~(\ref{7}) has the general solution
\begin{equation}
w(W,k)=\frac{C}{V_{f}}We^{-H(W,k)},
\end{equation}%
while for $v$ we obtain
\begin{equation}
v(W,k)=\frac{C}{V_{f}}D\left[ C^{-1}e^{H(W,k)}\right] We^{-H(W,k)},
\end{equation}

Therefore, by using the Chiellini Lemma, we have obtained the following
general solution for the general reaction-diffusion equation Eq.~(\ref{2}):

\textbf{Theorem 1.} \textit{If the diffusion function $D(c)$ and the
reaction function $Q(c)$ in a general one-dimensional reaction-diffusion
equation satisfy the condition given by Eq.~(\ref{cond}), then the equation
admits exact travelling wave solutions, expressed in parametric form as
\begin{equation}  \label{xi}
\xi (W,k)-\xi _{0}\left(W_0,k\right)=\frac{1}{V_{f}}\int _{W_0}^W{\frac{D%
\left[ C^{-1}e^{H(\psi ,k)}\right] }{1+\psi+k\psi^{2}}d\psi},
\end{equation}%
\begin{equation}  \label{c}
c(W,k)=C^{-1}e^{H(W,k)},
\end{equation}
 where $\xi _{0}$ is an arbitrary constant of integration, and we have
taken $W $ as a parameter.}

The constant $k$ can be determined once the explicit functional dependence of $D$ and $Q$ on $c$ is
known as
\begin{equation}
k=\frac{D\left( c_{0}\right) Q\left( c_{0}\right)}{V_{f}^{2}c_{0}},
\end{equation}
while the integration constant $C$ is determined as
\begin{equation}
C=\frac{e^{H\left( W_{0},k\right) }}{c_{0}}.
\end{equation}

For $W=W_{0}$, $\xi \left( W_{0},k\right) =\xi _{0}\left( W_{0},k\right) =0$%
, a condition which determines the integration constant $\xi _{0}\left(
W_{0},k\right) $ as $\xi _{0}\left( W_{0},k\right) =0$.

\subsection{Travelling wave solutions for  $D(c)=\mathrm{constant}$ and $Q(c)\propto c$}

As a first example of the application of \textbf{Theorem 1} we consider the
case when the diffusion and the reaction functions are given by $D(c)=D_{0}=%
\mathrm{constant}$ and $Q(c)=\rho c$, $\rho =\mathrm{constant}$. Therefore
the equation describing the travelling wave propagation in the system takes
the form
\begin{equation}
\frac{d^{2}c}{d\xi ^{2}}+\frac{V_{f}}{D_{0}}\frac{dc}{d\xi }+\frac{\rho }{%
D_{0}}c=0.  \label{ex1}
\end{equation}%
The integrability condition given by Eq.~(\ref{cond}) fixes the constant $k$
as $k=\rho D_{0}/V_{f}^{2}>1/4$. Then from Eq.~(\ref{xi}) we obtain the
 function $\xi (W)$ as
\begin{equation}
\xi (W)-\xi _{0}=-\frac{2D_{0}}{\Delta }\tanh ^{-1}\left( \frac{2D_{0}\rho
W+V_{f}^{2}}{V_{f}\Delta }\right) ,\frac{\rho D_{0}}{V_{f}^{2}}>\frac{1}{4},
\end{equation}%
giving
\begin{equation}
W\left( \xi \right) =-\frac{V_{f}}{2D_{0}\rho }\left\{ V_{f}+\Delta \tanh %
\left[ \frac{\Delta }{2D_{0}}\left( \xi -\xi _{0}\right) \right] \right\}
,\frac{\rho D_{0}}{V_{f}^{2}}>\frac{1}{4},
\end{equation}%
where we have denoted $\Delta =\sqrt{V_{f}^{2}-4D_{0}\rho }$. The general
solution of Eq.~(\ref{ex1}) is given by
\begin{equation}
c(\xi )=\sum_{n=+,-}C_{n}e^{-\frac{1}{2D_{0}}\left( V_{f}+n\Delta \right)
\xi },\frac{\rho D_{0}}{V_{f}^{2}}>\frac{1}{4},  \label{dd}
\end{equation}%
where $C_{n}$ , $n=+,-$ are arbitrary constants of integration. Note that
the general solution of Eq.~(\ref{ex1}) as given by Eq.~(\ref{dd}) can be
obtained directly from Eq.~(\ref{c}), which is a linear second order homogeneous differential equation.

\subsection{Travelling wave solutions for $D(c)\propto \left(1-c/c_{max}\right)^{-1}$, $Q(c)\propto  c\left(1-c/c_{max}\right)$}

As a second example of application of \textbf{Theorem 1} we consider that
the diffusion and the reaction functions are given by
\begin{equation}
D(c)=\frac{D_0}{1-c/c_{max}},Q(c)=\rho c\left(1-\frac{c}{c_{max}}%
\right),D_0,\rho ,c_{max} =\mathrm{constant}.
\end{equation}

The travelling wave equation for the reaction-diffusion system with these
forms of $D$ and $Q$ is given by
\begin{equation}
\frac{d^{2}c}{d\xi ^{2}}+\frac{1}{c_{max}\left(1-c/c_{max}\right)}\left(
\frac{dc}{d\xi }\right) ^{2}+\frac{V_{f}}{D_{0}}\left( 1-\frac{c}{c_{max}}%
\right) \frac{dc}{d\xi }+\frac{\rho }{D_{0}}c\left( 1-\frac{c}{c_{max}}%
\right) ^{2}=0.  \label{33}
\end{equation}

Eq.~(\ref{33}) represents the travelling wave form for a generalized Fisher
equation given by
\begin{equation}
\frac{\partial c(x,t)}{\partial t}=\frac{\partial }{\partial x}\left[ \frac{%
D_{0}}{1-c(x,t)/c_{max}}\frac{\partial c(x,t)}{\partial x}\right] +\rho
c(x,t)\left[ 1-\frac{c(x,t)}{c_{max}}\right] .
\end{equation}

Since the functions $D(c)$ and $Q(c)$ satisfy the integrability condition
given by Eq.~(\ref{cond}), with $D_{0}\rho =kV_{f}^{2}$, the general
solution of Eq.~(\ref{33}) can be obtained in an exact parametric form.

\subsubsection{The case $k=1/4$}

We
consider first the case $k=1/4$. Then the general solution of Eq.~(\ref{33})
is given in parametric form as
\begin{equation}
\xi (W)-\xi _{0}=\frac{4CD_{0}c_{max}}{V_{f}}\int_{W_{0}}^{W}\frac{\,d\psi }{%
\left( \psi +2\right) \left[ Cc_{max}(\psi +2)-e^{\frac{2}{\psi +2}}\psi %
\right] },V_{f}=2\sqrt{D_{0}\rho },
\end{equation}%
\begin{equation}
c(W)=C^{-1}\frac{W}{W+2}\exp \left( \frac{2}{W+2}\right) ,
\end{equation}%
where $C$ is an arbitrary constant of integration.

\subsubsection{The case $k\neq 1/4$}

For $k\neq 1/4$, we
obtain
\begin{eqnarray}
\xi (W,k)-\xi _{0}&=&\frac{D_{0}Cc_{max}}{V_{f}}\int_{W_{0}}^{W}{\left[
Cc_{max}\left( k\psi ^{2}+\psi +1\right) -\frac{\psi \sqrt{k\psi ^{2}+\psi +1}}{e^{%
\frac{\tan ^{-1}\left( \frac{2k\psi +1}{\sqrt{4k-1}}\right) }{\sqrt{4k-1}}}}%
\right] ^{-1}d\psi },\nonumber\\
&&k=\frac{\rho D_{0}}{V_{f}^{2}}> 1/4,
\end{eqnarray}%
\begin{equation}
c(W,k)=\frac{W}{C\sqrt{kW^{2}+W+1}}\exp \left( -\frac{1}{\sqrt{4k-1}}%
\tan ^{-1}\frac{1+2kW}{\sqrt{4k-1}}\right) ,k=\frac{\rho D_{0}}{V_{f}^{2}}%
> 1/4,
\end{equation}%
\begin{eqnarray}
\xi (W,k)-\xi _{0}&=&\frac{D_{0}Cc_{max}}{V_{f}}\int_{W_{0}}^{W}{\left[
Cc_{max}\left( k\psi ^{2}+\psi +1\right) -\frac{\psi \sqrt{k\psi ^{2}+\psi +1}}{e^{%
\frac{{\rm tanh} ^{-1}\left( \frac{2k\psi +1}{\sqrt{1-4k}}\right) }{\sqrt{1-4k}}}}%
\right] ^{-1}d\psi },\nonumber\\
&&k=\frac{\rho D_{0}}{V_{f}^{2}}< 1/4,
\end{eqnarray}%
\begin{equation}
c(W,k)=\frac{W}{C\sqrt{kW^{2}+W+1}}\exp \left( -\frac{1}{\sqrt{1-4k}}%
\tanh ^{-1}\frac{1+2kW}{\sqrt{1-4k}}\right) ,k=\frac{\rho D_{0}}{V_{f}^{2}}%
< 1/4,
\end{equation}
where the constant $k$ is determined by the model parameters $V_{f}$, $D_{0}$
and $\rho $.

\subsection{Travelling wave solutions of the first generalized Fisher-Kolmogorov equation}

Exact travelling wave solutions of more general equations of the form
\begin{equation}
\frac{\partial c(x,t)}{\partial t}=\frac{\partial }{\partial x}\left\{ \frac{%
D_{0}}{\left[ 1-c(x,t)/c_{max}\right] ^{\alpha }}\frac{\partial c(x,t)}{%
\partial x}\right\} +\rho c(x,t)\left[ 1-\frac{c(x,t)}{c_{max}}\right]
^{\alpha },  \label{mod0}
\end{equation}%
where $D_{0}$, $\alpha $ and $\rho $ are arbitrary constants, are given, in
a parametric form, by the following equations,
\begin{equation}
\xi (W,k)-\xi _{0}\left( W_{0},k\right) =\frac{D_{0}}{V_{f}}\int_{W_{0}}^{W}%
\frac{{d\psi }}{\left[ 1-C^{-1}c_{max}^{-1}e^{H(\psi ,k)}\right] ^{\alpha
}\left( 1+\psi +k\psi ^{2}\right) },  \label{mod1}
\end{equation}%
and
\begin{equation}
c(W,k)=C^{-1}e^{H(W,k)},  \label{mod2}
\end{equation}%
respectively, with $k=\rho D_{0}/V_{f}^{2}$. In the following we will call
Eq.~(\ref{mod0}) {\it the first generalized Fisher-Kolmogorov equation}.

\section{Further exact travelling wave solutions of the general
reaction-diffusion equation}\label{sect4}

In the following we  introduce first a new variable $\theta (c)$ through the Lemke transformation, defined as
\begin{equation}
w=-\frac{1}{V_{f}}\frac{d\ln \left\vert \theta \right\vert }{dc}.
\end{equation}%
Therefore Eq.~(\ref{5}) becomes
\begin{equation}
\frac{d^{2}\theta }{dc^{2}}=\frac{1}{V_{f}^{2}}D(c)Q(c)\frac{1}{\theta ^{2}}%
\left( \frac{d\theta }{dc}\right) ^{3}.  \label{25}
\end{equation}%
By taking into account the differential identity
\begin{equation}
\frac{d^{2}c}{d\theta
^{2}}=-\left( \frac{d^{2}\theta}{dc^{2}}\right) \left( \frac{d\theta }{dc}\right) ^{-3},
\end{equation}
Eq.~(\ref{25}) takes the form \cite{kamke, Lemke}
\begin{equation}
\theta ^{2}\frac{d^{2}c}{d\theta ^{2}}+\frac{1}{V_{f}^{2}}D(c)Q(c)=0.
\label{26}
\end{equation}

Eq. (\ref{26}) must be integrated with the initial conditions $c\left(
\theta _{0}\right) =c_{0}$ and
\begin{equation}
\left.\frac{dc}{d\theta }\right |_{\theta =\theta _{0}}=-\frac{D\left(
c_{0}\right) c_{0}^{\prime }}{V_{f}\theta _{0}}
\end{equation}

Therefore we have obtained the following

\textbf{Theorem 2}. \textit{If a solution $c=c(\theta )$ of Eq.~(\ref{26}) is known, then
the general reaction-diffusion equation Eq.~(\ref{3}) admits travelling wave
solutions, given in parametric form as
\begin{equation}
c=c(\theta ),
\end{equation}
\begin{equation}
\xi -\xi _{0}=-\frac{1}{V_{f}}\int_{\theta _{0}}^{\theta }\psi {%
^{-1}D[c(\psi )]d\psi },
\end{equation}%
where $\xi _{0}$ is an arbitrary constant of integration.}

If the solution of
Eq.~(\ref{26}) is obtained in a parametric form, with parameter $\tau $,
then the general travelling wave solution of the reaction - diffusion
equation Eq.~(\ref{3}) is obtained as
\begin{equation}
c=c(\tau ),\theta =\theta (\tau ),
\end{equation}
\begin{equation}
\xi (\tau )-\xi _{0}=-\frac{1}{V_{f}}%
\int_{\tau _{0}}^{\tau }\theta {^{-1}(\psi )D\left[ c(\psi )\right] \frac{%
d\theta \left( \psi \right) }{d\psi }d\psi }.
\end{equation}%

\section{Travelling wave solutions of the tumor growth equation}\label{new}

Depending on the functional form of the product of the diffusion and reaction functions $D(c)$ and $Q(c)$,  Eq.~(\ref{26}) can be integrated exactly in a number of cases, thus leading,  with the help of {\bf Theorem 2}, to exact travelling wave solutions of the general diffusion--reaction Eq.~(\ref{3}), which we  present in the following.

\subsection{Travelling wave solutions for $D(c)Q(c)={\rm constant}$}

In the case
the arbitrary function $D(c)Q(c)$ is a constant $\alpha $,
\begin{equation}  \label{d1}
D(c)Q(c)=\alpha =\mathrm{constant},
\end{equation}
subsequently Eq.~(\ref{26}) has the general solution
\begin{equation}  \label{c1}
c(\theta )=C_{1}+C_{2}\theta +\frac{\alpha \ln |\theta |}{V_{f}^{2}},
\end{equation}%
where $C_{1}$ and $C_{2}$ are arbitrary constants of integration. Hence for $%
w$ we obtain
\begin{equation}
w\left( \theta \right) =-\frac{V_{f}}{\alpha +C_{2}\theta V_{f}^{2}}.
\label{ww}
\end{equation}

Eq.~(\ref{ww}) identically satisfies Eq.~(\ref{5}), and therefore we can
take the arbitrary integration constant $C_1$ as zero, $C_1=0$. Using the
transformations $v\left( \theta \right) =D\left( \theta \right) w\left(
\theta \right) $, $dc/d\xi =1/v\left( \theta \right) $, and with the help of
Eq.~(\ref{ww}), we obtain the following expression for $\xi $, giving,
together with Eq.~(\ref{c1}), the general solution of the general
reaction-diffusion Eq.~(\ref{2}) with the coefficients $D(c)$ and $Q(c)$
satisfying the condition (\ref{d1}) as,
\begin{equation}
\xi (\theta )-\xi _{1}=-\frac{1}{V_{f}}\int_{\theta _{0}}^{\theta }\psi
^{-1}D\left[( C_{2}\psi +\frac{\alpha \ln |\psi |}{V_{f}^{2}}\right) {d\psi }%
,  \label{xi1}
\end{equation}%
where $\xi _{1}$ is an arbitrary constant of integration, and we have taken $%
\theta $ as a parameter. Eqs.~(\ref{c1}) and (\ref{xi1}) give the general
solution of the general reaction-diffusion equation Eq.~(\ref{3}) with
diffusion and reaction functions satisfying the condition (\ref{d1}). In
order to determine the integration constants $C_{1}$ and $C_{2}$, we use the
initial conditions that give
\begin{equation}
c(\theta _{0})=C_{2}\theta _{0}+\frac{\alpha \ln |\theta _{0}|}{V_{f}^{2}}%
=c_{0},
\end{equation}
and
\begin{equation}
\left. \frac{dc}{d\theta }\right |_{\theta =\theta _{0}}=C_{2}+\frac{\alpha
}{V_{f}^{2}\theta _{0}}=-\frac{D\left( c_{0}\right) c_{0}^{\prime }}{%
V_{f}\theta _{0}},
\end{equation}%
giving
\begin{equation}
C_{2}=-\frac{1}{V_{f}\theta _{0}}\left[ \frac{\alpha }{V_{f}}+D\left(
c_{0}\right) c_{0}^{\prime }\right] ,
\end{equation}%
and
\begin{equation}  \label{61}
c_{0}+\frac{1}{V_{f}}\left[ \frac{\alpha }{V_{f}}+D\left( c_{0}\right)
c_{0}^{\prime }\right] -\frac{\alpha \ln |\theta _{0}|}{V_{f}^{2}}=0,
\end{equation}%
respectively. Eqs.~(\ref{61}) determines the value of $\theta _0$ as a
function of the initial conditions $\left(c_0,c_0^{\prime }\right)$ and the
free parameters $\left\{\alpha, V_f,D\left(c_0\right)\right\}$.

\subsection{Travelling wave solutions with a linear dependence of $D(c)Q(c)$}

Next, in order to obtain another exact general solution of Eq.~(\ref{26}),
we assume that the diffusion function $D\left( c\right) $ and the reaction
function $Q\left( c\right) $ satisfy the condition
\begin{equation}
D\left( c\right) Q\left( c\right) =\beta c+\alpha ,  \label{dc}
\end{equation}%
where $\beta $ is an arbitrary constant. By inserting Eq.~(\ref{dc}) into
Eq.~(\ref{26}), the latter equation becomes
\begin{equation}
\theta ^{2}\frac{d^{2}c}{d\theta ^{2}}+\frac{\beta c+\alpha }{V_{f}^{2}}=0.
\label{dc1}
\end{equation}

Eq.~(\ref{dc1}) can be integrated to yield
\begin{equation}  \label{c2}
c\left( \theta \right) =C_{3}\theta ^{m_{+}}+C_{4}\theta ^{m_{-}}-\frac{%
\alpha }{\beta },
\end{equation}
where $C_{3}$ and $C_{4}$ are arbitrary constants of integration, and
\begin{equation}
2m_{\pm }=1\pm \sqrt{1-4\frac{\beta }{V_{f}^{2}}}.
\end{equation}

The function $v\left( \theta
\right) $ is given by
\begin{equation}
v\left( \theta \right) =-\frac{D\left( c\right) }{V_{f}\left(
C_{3}m_{+}\theta ^{m_{+}}+C_{4}m_{-}\theta ^{m_{-}}\right) }.
\end{equation}

Thus we obtain
\begin{equation}  \label{xi2}
\xi (\theta )-\xi _{2}=-\frac{1}{V_{f}}\int _{\theta _0}^{\theta } \psi
^{-1}D\left( C_{3}\psi ^{m_{+}}+C_{4}\psi ^{m_{-}}-\frac{\alpha }{\beta }%
\right) {d\psi },
\end{equation}%
where $\xi _{2}$ is an arbitrary constant of integration, and we have taken $%
\theta $ as a parameter. Thus the general solution of the reaction -
diffusion equation Eq.~(\ref{3}) with diffusion and reaction functions
satisfying the condition given by Eq.~(\ref{dc}) is given, in a parametric
form, by Eqs.~(\ref{c2}) and Eq.~(\ref{xi2}), respectively.

The numerical values of the integration constants $C_{3}$ and $C_{4}$ are
determined from the initial conditions as
\begin{equation}
C_3=\frac{\theta _0^{-\frac{1}{2} \sqrt{1-\frac{4 \beta }{V_f^2}}-\frac{1}{2}}
   \left[\left(\alpha +\beta  c_0\right) \left(\sqrt{1-\frac{4 \beta }{V_f^2}}-1\right)-2 \beta D\left(c_0\right)c_0'/V_f
   \right]}{2 \beta  \sqrt{1-\frac{4 \beta }{V_f^2}}},
\end{equation}
and
\begin{equation}
C_4=\frac{\theta _0^{\frac{1}{2} \sqrt{1-\frac{4 \beta }{V_f^2}}-\frac{1}{2}}
   \left[\left(\alpha +\beta  c_0\right) \left(\sqrt{1-\frac{4 \beta
   }{V_f^2}}+1\right)+2 \beta
   D\left(c_0\right)c_0'/V_f\right]}{2 \beta  \sqrt{1-\frac{4 \beta }{V_f^2}}},
\end{equation}%
respectively.

If the arbitrary constant $\alpha $ vanishes, then we obtain the condition $%
D(c)Q(c)=\beta c$ given by Eq.~(\ref{cond}), thus we regain the solution
obtained in the previous Section by using the Chiellini lemma.

\subsection{Travelling wave solutions with $D(c)Q(c)\propto 1/c$}

If the diffusion and the reaction functions $D(c)$ and $Q(c)$ satisfy the
relation
\begin{equation}  \label{d3}
D(c)Q(c)=\frac{KV_f^2}{c},
\end{equation}
where $K$ is a constant, Eq.~(\ref{26}) takes the form of an Emden-Fowler
equation \cite{Pol},
\begin{equation}  \label{50}
\frac{d^2c}{d\theta ^2}+K\theta ^{-2}c^{-1}=0.
\end{equation}
Eq.~(\ref{50}) has the exact solution, given in parametric form,
\begin{equation}  \label{c3}
\theta (\tau )=C_5\left[\left(\sqrt{\pi }/2\right)\mathrm{erf}(\tau)+C_6%
\right]^{-1},
\end{equation}
\begin{equation}
c(\tau )=\sqrt{\frac{K}{2}}\frac{d}{d\tau}\ln \left[\left(\sqrt{%
\pi }/2\right)\mathrm{erf}(\tau)+C_6\right],
\end{equation}
where $C_5$ and $C_6$ are arbitrary constants of integration, and $\mathrm{%
erf}(z)=\left(2/\sqrt{\pi}\right)\int _0^z{e^{-t^2}dt}$ is the error
function, representing the integral of the Gaussian distribution. For the
parametric dependence of $\xi $ we obtain
\begin{equation}  \label{xi3}
\xi (\tau)-\xi _0=\frac{2}{V_f}\int_{\tau _0}^{\tau }{\frac{ e^{-\psi
^2}D[c(\psi)]}{2 C_6+\sqrt{\pi } \text{erf}(\psi )}d\psi}.
\end{equation}
Eqs.~(\ref{c3}) and (\ref{xi3}) give the general solution of the reaction -
diffusion equation with diffusion and reaction functions satisfying the
condition (\ref{d3}).

\subsection{Travelling waves solutions with a power law dependence of $D(c)Q(c)$}

If $D(c)$ and $Q(c)$ satisfy the relation
\begin{equation}
\frac{1}{V_{f}^{2}}D(c)Q(c)=\frac{2(m+1)}{(m+3)^{2}}c-Ac^{m},m\neq -1,m\neq
-3,  \label{d4}
\end{equation}%
then the basic equation determining the travelling wave solutions of the
corresponding general reaction-diffusion equation is
\begin{equation}  \label{mods0}
\frac{d^{2}c}{d\theta ^{2}}+\frac{2(m+1)}{(m+3)^{2}}\theta ^{-2}c-A\theta
^{-2}c^{m}=0,
\end{equation}%
and it has the exact parametric solution given by \cite{Pol}
\begin{equation}
\theta \left( \tau \right) =C_{7}\left( \int ^{\tau}{\frac{d\psi }{\sqrt{%
1+\psi ^{m+1}}}}+C_{8}\right) ^{\left( m+3\right) /(m-1)},m\neq -1,m\neq -3,m\neq 1,
\end{equation}
\begin{equation}\label{c4}
c\left( \tau\right) =b\tau \left( \int ^{\tau } \frac{d\psi }{\sqrt{1+\psi ^{m+1}}}%
+C_{8}\right) ^{2/(m-1)},
m\neq -1,m\neq -3,m\neq 1,
\end{equation}%
where
\begin{equation}
b=\left[ \frac{\left( m-1\right) ^{2}\left( m+1\right) }{2A\left( m+3\right)
^{2}}\right] ^{1/\left( m-1\right) },
\end{equation}%
and $C_{7}$ and $C_{8}$ are arbitrary constants of integration. The
parametric dependence of $\xi $ is obtained as
\begin{equation}
\xi (\tau )-\xi _{0}=-\frac{m+3}{(m-1)V_{f}}\int ^{\tau } {\frac{D[c(\psi
)]d\psi }{\sqrt{\psi ^{m+1}+1}\left[ C_{8}+\psi \,_{2}F_{1}\left( \frac{1}{2}%
,\frac{1}{m+1};1+\frac{1}{m+1};-\psi ^{m+1}\right) \right] }},  \label{xi4}
\end{equation}%
where $_{2}F_{1}[a,b;c,z]=\sum_{k=0}^{\infty }{\left[ (a)_{k}(b)_{k}/(c)_{k}%
\right] z^{k}/k!}$ is the hypergeometric function. Eqs.~(\ref{c4}) and (\ref%
{xi4}) give the general solution of the general reaction-diffusion equation
with diffusion and reaction functions satisfying the condition (\ref{d4}).

\section{ Biological applications}\label{sect4a}

In the present Section we briefly point out some possibilities of biological
applications of the obtained results to model the growth of glioblastomas.
In order to obtain some numerical results we need to chose some values of
the parameters $D$ and $\rho $ that characterize the tumor dynamics. These
parameters can be computed observationally from as few as two pre-treatment
MRI observations \cite{9}, and current data from 29 tumors show a range of 6
-- 324 \textrm{mm}$^{2}$/year for $D$ and 1 -- 32 /year for $\rho $.
However, in the following, we describe the virtual tumor with parameter
values as follows: $D=1.43\;\mathrm{cm}^{2}/\mathrm{year}$, and $\rho =16.25/%
\mathrm{year}$, which serve as representative means of published ranges \cite%
{9}. The quantity $c$ represent the number of tumor cells within the volume $%
V$. For the sake of simplicity we fix the initial concentration of tumoral
cells as $c(0)=c_{0}=1000\mathrm{\;cells}/$cm$^{3}$ \cite{incell}, and we take $\left(
dc/d\xi \right) |_{\xi =0}\approx 7\times 10^{8}$ cells/cm$^4$. To estimate $c_{max}$, we assume
that the volume of a typical cell is 1200 $\mu$m$^3$, as it is for EMT6/Ro
tumor cells \cite{33}. Assuming that half the volume of the spheroid is made
up of tumor cells, the maximum density is $c_{max}= 4.2 \times 10^8$ cells/cm%
$^3$ \cite{1}. This tumor model is only applicable for tumors having a
volume greater than $1\;\mathrm{mm}^3$ \cite{3,1}.

In the following we will consider two tumor growth toy models, described by
the travelling wave solutions of the generalized Fisher--Kolmogorov equations
Eq.~(\ref{mod0}), and by Eq.~(\ref{mods0}), respectively.

\subsection{Tumor growth in the first generalized Fisher--Kolmogorov
equation model}

We will investigate first the properties of the travelling wave model for
tumor growth in the first generalized Fisher--Kolmogorov equation,
\begin{equation}\label{FK1}
\frac{\partial c(x,t)}{\partial t}=\frac{\partial }{\partial x}\left\{ \frac{%
D_{0}}{\left[ 1-c(x,t)/c_{max}\right] ^{\alpha }}\frac{\partial c(x,t)}{%
\partial x}\right\} +\rho c(x,t)\left[ 1-\frac{c(x,t)}{c_{max}}\right]
^{\alpha },
\end{equation}%
whose solutions are given by Eqs.~(\ref{mod1}) and (\ref{mod2}),
respectively. In the present
Section for $D_{0}$ we adopt the numerical value $D_{0}=D=1.43\;\mathrm{cm}%
^{2}/\mathrm{year}$.

\subsubsection{ The case $k=1/4$}

We will begin our analysis of the travelling wave solutions in the first
generalized Fisher-Kolmogorov model by considering the case $k=1/4$. The
velocity of the travelling front is given in this model by
\begin{equation}
V_{f}=2\sqrt{D_{0}\rho }=9.6411\;\mathrm{cm/year}.
\end{equation}%
Then the general travelling wave solution of the first generalized Fisher-Kolmogorov equation
is given in parametric form as
\begin{equation}
\xi (W)-\xi _{0}\left( W_{0}\right) =\frac{4D_{0}}{V_{f}}\int_{W_{0}}^{W}%
\frac{{d\psi }}{\left[ 1-C^{-1}c_{max}^{-1}\frac{\psi }{\psi +2}\exp \left(
\frac{2}{\psi +2}\right) \right] ^{\alpha }\left( 2+\psi \right) ^{2}},
\label{78}
\end{equation}%
and
\begin{equation}
c(W)=C^{-1}\frac{W}{W+2}\exp \left( \frac{2}{W+2}\right) ,
\end{equation}%
respectively. By performing a series expansion of the integrand in Eq.~(\ref%
{78}) we obtain
\begin{eqnarray}
&&\left\{ \left[ 1-C^{-1}c_{max}^{-1}\frac{\psi }{\psi +2}\exp \left( \frac{2}{%
\psi +2}\right) \right] ^{\alpha }\left( 2+\psi \right) ^{2}\right\} ^{-1}
\approx \nonumber\\
&&\frac{1}{4}+\frac{1}{8}   \left(\frac{e \alpha }{K}-2\right)\psi+
\frac{ \left[e^2 \alpha  (\alpha +1)+6 K^2-8 e \alpha  K\right]}{32 K^2}\psi ^2+\nonumber\\
&&\frac{ \left[e^3 \alpha  \left(\alpha ^2+3 \alpha +2\right)-24 K^3+63 e \alpha  K^2-18
   e^2 \alpha  (\alpha +1) K\right]}{192 K^3}\psi ^3+O\left(\psi ^4\right),
\end{eqnarray}%
where $K=Cc_{max}$, giving
\begin{eqnarray}
\frac{V_f}{4D_0}\xi (W) &\approx &\frac{1}{4}\left( W-W_{0}\right) +\frac{1}{%
8}\left( \frac{e\alpha }{2K}-1\right) \left( W^{2}-W_{0}^{2}\right) +\nonumber\\
&&\frac{%
\left[ e^{2}\alpha \left( \alpha +1\right) +2K\left( 3K-4e\alpha \right) %
\right] }{96K^{2}}\left( W^{3}-W_{0}^{3}\right) +  \notag  \label{approx1} \\
&&\frac{\left[ \alpha e^{3}\left( \alpha ^{2}+3\alpha +2\right)
+3K^{2}\left( 21e\alpha -8K\right) -18e^{2}\alpha \left( \alpha +1\right) K%
\right] }{768K^{3}}\times \nonumber\\
&&\left( W^{4}-W_{0}^{4}\right).
\end{eqnarray}

The cell concentration $c(W)$ can be obtained approximately as
\begin{eqnarray}  \label{approx2}
c(W) &\approx &\frac{e^{\frac{2}{W_{0}+2}}W_{0}}{C(W_{0}+2)}+\frac{4e^{\frac{%
2}{W_{0}+2}}}{C(W_{0}+2)^{3}}(W-W_{0})-\frac{2e^{\frac{2}{W_{0}+2}}(3W_{0}+8)%
}{C(W_{0}+2)^{5}}(W-W_{0})^{2}+  \notag \\
&&\frac{8e^{\frac{2}{W_{0}+2}}\left( 3W_{0}^{2}+16W_{0}+21\right) }{%
3C(W_{0}+2)^{7}}(W-W_{0})^{3}-\nonumber\\
&&\frac{2e^{\frac{2}{W_{0}+2}}\left(
15W_{0}^{3}+120W_{0}^{2}+315W_{0}+272\right) }{3C(W_{0}+2)^{9}}(W-W_{0})^{4}.
\end{eqnarray}

Eqs.~(\ref{approx1}) and (\ref{approx2}) give an approximate parametric
representation of the travelling wave solution of the generalized
Fisher-Kolmogorov equation Eq.~(\ref{mod0}) for $k=1/4$. In the first order
of approximation,
\begin{equation}
c(\xi )\approx \frac{e^{\frac{2}{W_{0}+2}}W_{0}}{C(W_{0}+2)}+\frac{4V_fe^{%
\frac{2}{W_{0}+2}}}{CD_0(W_{0}+2)^{3}}\xi .
\end{equation}

In the limit $W\rightarrow \infty $ we have
\begin{equation}
\lim_{W\rightarrow \infty
}c\left( W\right) =C^{-1}=c_{\max }^{(0)^{{}}}=\mathrm{constant}, k=\frac{1}{4}.
\end{equation}

\begin{figure}
   \centering
  \includegraphics[width=8cm]{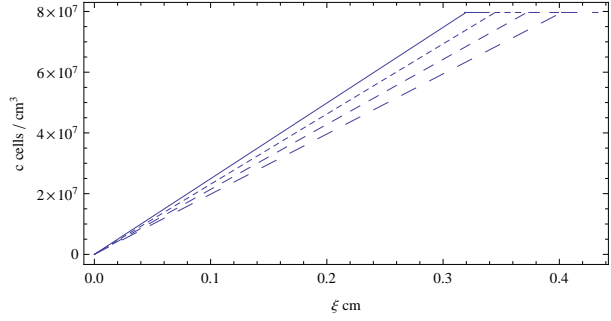}
  \caption{Variation of the cell number density $c$ as a function of $\xi $ for the first generalized Fisher-Kolmogorov model with $V_f=2\sqrt{D_0\rho }=9.64$ cm/year, for different values of $\alpha $: $\alpha =1/2$ (solid curve), $\alpha =1$ (dotted curve), $\alpha =3/2$ (short dashed curve) and $\alpha =2$ (dashed curve), respectively. The initial values of the cell concentration and its derivative are $c(0)=1000$ cells/cm$^3$ and $\left.(dc/d\xi)\right |_{\xi =0}=7.3\times 10^8$ cells/cm$^4$.}\label{fig1}
  \end{figure}

The variation of the cell number density $c$ as a function of $\xi $ is represented in Fig.~\ref{fig1}.  As one can see from the figure, the cell number density increases linearly with $\xi$, and reaches a constant value at a finite $\xi$.

\subsubsection{The case $k\neq 1/4$}

For $k> 1/4$, the general solution of the first generalized Fisher--Kolmogorov
equation is given by
\begin{equation}
\xi (W,k)-\xi _{0}\left( W_{0},k\right) =\frac{D_{0}}{V_{f}}\int_{W_{0}}^{W}{%
\frac{d\psi }{\left[ 1-\frac{\psi e^{-\frac{\tan ^{-1}\left(
\frac{2k\psi +1}{\sqrt{4k-1}}\right) }{\sqrt{4k-1}}}}{K\sqrt{1+\psi +k\psi ^{2}%
}}\right] ^{\alpha }\left( 1+\psi +k\psi ^{2}\right) }},  \label{84}
\end{equation}%
and
\begin{equation}
c(W,k)=C^{-1}\frac{We^{-\frac{\tan ^{-1}\left( \frac{2kW+1}{\sqrt{4k-1}}%
\right) }{\sqrt{4k-1}}}}{\sqrt{1+W+kW^{2}}},
\end{equation}%
where $k=\rho D_{0}/V_{f}^{2}>1/4$. In the second order of approximation the
integrand in Eq.~(\ref{84}) becomes
\begin{eqnarray}
&&\frac{\left[ 1-K^{-1}\psi e^{-\frac{\tan ^{-1}\left( \frac{%
2k\psi +1}{\sqrt{4k-1}}\right) }{\sqrt{4k-1}}}/\sqrt{1+\psi +k\psi ^{2}}%
\right] ^{-\alpha }}{1+\psi +k\psi ^{2}} \approx 1+
   \left[\frac{\alpha  e^{-\frac{\tan ^{-1}\left(\frac{1}{\sqrt{4
   k-1}}\right)}{\sqrt{4 k-1}}}}{K}-1\right]\psi+\nonumber\\
&&\frac{ e^{-\frac{2 \tan ^{-1}\left(\frac{1}{\sqrt{4 k-1}}\right)}{\sqrt{4
   k-1}}} \left[\alpha  (\alpha +1)-2 (k-1) K^2 e^{\frac{2 \tan
   ^{-1}\left(\frac{1}{\sqrt{4 k-1}}\right)}{\sqrt{4 k-1}}}-4 \alpha  K e^{\frac{\tan
   ^{-1}\left(\frac{1}{\sqrt{4 k-1}}\right)}{\sqrt{4 k-1}}}\right]}{2 K^2}\psi ^2,\nonumber\\
\end{eqnarray}%
giving
\begin{eqnarray}
&&\frac{V_{f}}{D_{0}}\left[ \xi (W,k)-\xi _{0}\left( W_{0},k\right) \right]
\approx \left( W-W_{0}\right) +\frac{1}{2}\left[ \frac{\alpha e^{-\frac{%
\tan ^{-1}\left( \frac{1}{\sqrt{4k-1}}\right) }{\sqrt{4k-1}}}}{K}-1\right]
\left( W^{2}-W_{0}^{2}\right) + \nonumber\\
&&\frac{ e^{-\frac{2 \tan ^{-1}\left(\frac{1}{\sqrt{4 k-1}}\right)}{\sqrt{4
   k-1}}} \left[\alpha  (\alpha +1)-2 (k-1) K^2 e^{\frac{2 \tan
   ^{-1}\left(\frac{1}{\sqrt{4 k-1}}\right)}{\sqrt{4 k-1}}}-4 \alpha  K e^{\frac{\tan
   ^{-1}\left(\frac{1}{\sqrt{4 k-1}}\right)}{\sqrt{4 k-1}}}\right]}{6 K^2}\times \nonumber\\
&&\left(W^3-W_0^3\right)+....
\end{eqnarray}

In the same order of approximation for the cell density we obtain
\begin{eqnarray}
Cc(W,k) &\approx &\frac{W_{0}e^{-\frac{\tan ^{-1}\left( \frac{2kW_{0}+1}{%
\sqrt{4k-1}}\right) }{\sqrt{4k-1}}}}{\sqrt{kW_{0}^{2}+W_{0}+1}}+\frac{e^{-%
\frac{\tan ^{-1}\left( \frac{2kW_{0}+1}{\sqrt{4k-1}}\right) }{\sqrt{4k-1}}}}{%
\left( kW_{0}^{2}+W_{0}+1\right) ^{3/2}}(W-W_{0})-\nonumber\\
&&\frac{
(3kW_{0}+2)e^{-\frac{\tan ^{-1}\left( \frac{2kW_{0}+1}{\sqrt{4k-1}}\right) }{%
\sqrt{4k-1}}} }{2\left( kW_{0}^{2}+W_{0}+1\right) ^{5/2}}%
(W-W_{0})^{2}+  \notag \\
&&\frac{\left( 12k^{2}W_{0}^{2}+16kW_{0}-3k+6\right) e^{-\frac{\tan
^{-1}\left( \frac{2kW_{0}+1}{\sqrt{4k-1}}\right) }{\sqrt{4k-1}}}}{6\left(
kW_{0}^{2}+W_{0}+1\right) ^{7/2}}(W-W_{0})^{3}.
\end{eqnarray}

In the first order of approximation the cell density is given by
\begin{equation}
c(\xi)\approx \frac{W_{0}e^{-\frac{\tan ^{-1}\left( \frac{2kW_{0}+1}{\sqrt{%
4k-1}}\right) }{\sqrt{4k-1}}}}{C\sqrt{kW_{0}^{2}+W_{0}+1}}+\frac{V_{f}}{D_{0}%
}\frac{e^{-\frac{\tan ^{-1}\left( \frac{2kW_{0}+1}{\sqrt{4k-1}}\right) }{%
\sqrt{4k-1}}}}{C\left( kW_{0}^{2}+W_{0}+1\right) ^{3/2}}\xi (W,k),
\end{equation}
which shows that for small $\xi $ the cell number density is a linearly increasing function of $\xi $.

In the limit of large $W$ the cell number density tends to a constant value,
given by
\begin{equation}
\lim_{W\rightarrow \infty}c(W,k)=\frac{e^{-\frac{\pi }{2 \sqrt{4 k-1}}}}{C \sqrt{k}%
},k=\frac{\rho D_0}{V_f^2}>\frac{1}{4},
\end{equation}
and
\begin{equation}
\lim_{W\rightarrow \infty}c(W,k)=\frac{(-k)^{-\frac{1}{2 \sqrt{1-4 k}}} k^{\frac{1%
}{2} \left(\frac{1}{\sqrt{1-4 k}}-1\right)}}{C},k=\frac{\rho D_0}{V_f^2}<%
\frac{1}{4},
\end{equation}
respectively.

\begin{figure}
   \centering
  \includegraphics[width=8cm]{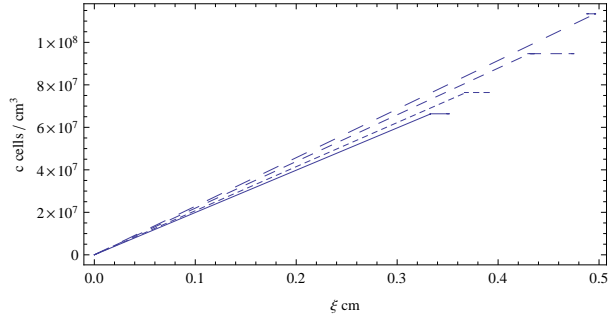}
  \caption{Variation of the cell number density $c$ as a function of $\xi $ for the first generalized Fisher-Kolmogorov model for $\alpha =3/2$ and for different values of $V_f$: $V_f=\sqrt{6D_0\rho }$ (solid curve), $V_f=\sqrt{4D_0\rho }$ (dotted curve), $V_f=\sqrt{2D_0\rho }$ (short dashed curve) and $V_f=\sqrt{D_0\rho }$ (dashed curve), respectively. The initial values of the cell concentration and its derivative are $c(0)=1000$ cells/cm$^3$ and $\left.(dc/d\xi)\right |_{\xi =0}=7\times 10^8$ cells/cm$^4$.}\label{fig2}
  \end{figure}

The variation of the cell number density $c$ with respect to $\xi $ is represented in Fig.~\ref{fig2}. The cell number linearly increases as a function of $\xi $, and reaches a maximum constant value, whose numerical value is dependent on the value of $V_f$.

  \subsubsection{Biological implications}

  One of the important predictions of the standard Fisher--Kolmogorov equation Eq.~(\ref{FK}) is that the velocity $v_{FK}$ of the detectable
tumor margin approximately satisfies the relation  $v_{FK}\approx2\sqrt{\rho D}$ \cite{8}. This result was obtained  from the observation that a  cell population with a dynamics determined by diffusion and growth alone expands at a constant velocity $v_{FK}$ for large times, thus expanding linearly for any given value of the (constant) diffusion coefficient, and $\rho $, respectively. An experimental study of the growth of low grade gliomas was performed in \cite{Man},   and it indeed confirmed that low-grade gliomas do grow both slowly and linearly. As for the tumor growth velocity, in the first 27 patients studied in \cite{Man}, the average velocity of the diameter is about 4 mm/year. On the other hand a much higher radial tumor velocity, of the order of  12 mm/year, was reported in a single rare patient with a glioblastoma that was followed with repeated MRIs for a year without intervening treatment \cite{8}.

In the  glioblastoma growth models described by the first generalized Fisher--Kolmogorov equation, the assumption of the concentration depending diffusion coefficient, and of a generalized reaction function do allow a wide range of velocities for the tumor front. Interestingly, these velocities are independent of the constant $\alpha $ in Eq.~(\ref{FK1}), and they depend only on the constant value $D_0$ of the diffusion function $D(c)$ for $c=0$, $D_0=D(0)$, on the proportionality coefficient $\rho $ in the reaction function, and on the arbitrary constant $k$. For $k=1/4$, and for the adopted values of $D_0$ and $\rho $, the velocity $V_f$ of the tumor front is of the order of $V_f\approx 10$ cm/year, rather different to the value of 12 mm/year adopted in \cite{8}. In this case we obtain for the velocity of the tumor growth front the same relation as for the approximate expression of the velocity in the standard Fisher--Kolmogorov equation, $V_f|_{k=1/4}=v_{FK}$.   On the other hand, a very wide range of tumor front velocities $V_f=\sqrt{\rho D_0/k}$ can be obtained for $k\neq 1/4$, {\it without any need of modifying the numerical values of the basic model parameters $D_0$ and $\rho $}. This shows that by adopting tumor growth models with concentration dependent  diffusion and reaction functions, a wide variety of experimental/clinical observations could be modelled, and the differences in the tumor growth observations could be attributed to the variations in the diffusion and reaction properties of the tumors.

All the considered travelling wave solutions of the first generalized Fisher - Kolmogorov equation show a linear expansion of the tumor size, who reaches suddenly the plateau phase, where the cell density becomes a constant. The tumor size evolves linearly in both the small cell density and high cell density phases, and therefore it is a general property of the present models. This behavior is consistent with the experimental results presented in \cite{Man}. As shown in Fig.~\ref{fig1}, for $k=1/4$, there is strong dependence of the cell concentration on the parameter $\alpha $, determining the diffusion and the reaction laws. On the other hand, for a fixed $\alpha $ and for $k\neq 1/4$, the modifications in $V_f$, due to its $k$-dependence, do affect significantly the maximum value of the cells in the plateau phase, as well the moment when this maximum is reached.

 \subsection{Tumor growth in the second generalized Fisher--Kolmogorov model}

We consider now the model in which the diffusion and reaction functions
satisfy the condition given by Eq.~(\ref{d4}). Moreover, we assume $D(c)=D_0=%
\mathrm{constant}$, which fixes the reaction function as
\begin{equation}
Q(c)=\frac{2m+1}{(m+3)^2}\frac{V_f^2}{D_0}c\left[1-A_m\left(\frac{c}{c_{max}}%
\right)^{m-1}\right],m\neq-1,m\neq -3,
\end{equation}
where $A_m=A(m+3)^2/(2m+1)$.
The corresponding reaction-diffusion equation describing tumor growth is
given by
\begin{eqnarray}  \label{95}
\frac{\partial c(x,t)}{\partial t}&=&D_0\frac{\partial ^2c(x,t)}{\partial x ^2}
+\frac{2m+1}{(m+3)^2}\frac{V_f^2}{D_0}c(x,t)\left[1-A_m\left(\frac{c(x,t)}{%
c_{max}}\right)^{m-1}\right], \nonumber\\
&&m\neq-1,m\neq -3.
\end{eqnarray}

We call Eq.~(\ref{95}) {\it the second generalized Fisher-Kolmogorov equation}.
Its travelling wave solution is given, in a parametric form, by Eqs.~(%
\ref{c4}) and (\ref{xi4}). In Eq.~(\ref{c4}) we take, without any loss of
generality, $C_7=1$. The constant $b$ is given by
\begin{equation}
b=\left[\frac{(m-1)^2(m+1)c_{max}^m}{2(m+3)^2A_m}\right]^{1/(m-1)},m\neq -3,m\neq -1,m\neq 1.
\end{equation}

The dependence of the cell number density on the parameter $\tau $ is obtained as
\begin{eqnarray}
c(\tau )&=&b\tau \Bigg[ C_{8}+\tau \,_{2}F_{1}\left( \frac{1}{2},\frac{1}{m+1}%
;1+\frac{1}{m+1};-\tau ^{m+1}\right) -\nonumber\\
&&\tau _{0}\,_{2}F_{1}\left( \frac{1}{2},%
\frac{1}{m+1};1+\frac{1}{m+1};-\tau _{0}^{m+1}\right) \Bigg] ^{2/(m-1)},
\label{97}
\end{eqnarray}%
while the dependence of $\xi $ on the parameter $\tau $ is given by
\begin{equation}
\xi (\tau )-\xi _{0}=-\frac{(m+3)D_{0}}{(m-1)V_{f}}\int_{\tau _{0}}^{\tau }{%
\frac{d\psi }{\sqrt{\psi ^{m+1}+1}\left[ C_{8}+\psi \,_{2}F_{1}\left( \frac{1%
}{2},\frac{1}{m+1};1+\frac{1}{m+1};-\psi ^{m+1}\right) \right] }},
\end{equation}%
where the integration constant $\xi _{0}$ can be taken as zero without any
loss of generality. The integral can be computed exactly to give
\begin{equation}
\xi (\tau )=-\frac{(m+3)D_{0}}{(m-1)V_{f}}\ln \frac{\left[ C_{8}+\tau
\,_{2}F_{1}\left( \frac{1}{2},\frac{1}{m+1};1+\frac{1}{m+1};-\tau
^{m+1}\right) \right] }{\left[ C_{8}+\tau _{0}\,_{2}F_{1}\left( \frac{1}{2},%
\frac{1}{m+1};1+\frac{1}{m+1};-\tau _{0}^{m+1}\right) \right] }.  \label{100}
\end{equation}

By taking into account that for $\tau =\tau _{0}$ we have $c\left( \tau
_{0}\right) =c_{0}$, from Eq.~(\ref{97}) we obtain the value of the
integration constant $C_{8}$ as
\begin{equation}
C_{8}=\left( \frac{c_{0}}{b\tau _{0}}\right) ^{(m-1)/2}.
\end{equation}

Therefore for $c(\tau )$ we obtain
\begin{eqnarray}
c(\tau )&=&b\tau e^{-[2/(m+3)](Vf/D_{0})\xi }
\Bigg\{ \left( \frac{c_{0}}{b\tau
_{0}}\right) ^{(m-1)/2}+\nonumber\\
&&\tau _{0}\,_{2}F_{1}\left( \frac{1}{2},\frac{1}{m+1}%
;1+\frac{1}{m+1};-\tau _{0}^{m+1}\right) \times \nonumber\\
&&\left[ 1-e^{[(m-1)/(m+3)](Vf/D_{0})%
\xi }\right] \Bigg\} ^{2/(m-1)}.
\end{eqnarray}

By estimating
\begin{equation}
w=-\frac{1}{V_{f}} \frac{1}{\theta }\frac{d\theta }{d\tau }\frac{d\tau }{dc},
\end{equation}
at $\tau =\tau _{0}$, and by using Eq.~(\ref{inw}), it follows that
the initial value $\tau _{0}$ of the parameter $\tau $ is obtained as a
solution of the algebraic equation
\begin{eqnarray}
\frac{(m+3) \left(\frac{c_0}{b \tau
   _0}\right)^{-1}}{b V_f \left[(m-1)
   \sqrt{\tau _0^{m+1}+1} \left(\frac{c_0}{b \tau _0}\right){}^{\frac{m-1}{2}}+2 \tau
   _0\right]}=\frac{%
bV_f}{c_{0}^{\prime }D_{0}}.
\end{eqnarray}

In the first order of approximation we have
\begin{eqnarray}
&&\tau \,_{2}F_{1}\left( \frac{1}{2},\frac{1}{m+1};1+\frac{1}{m+1};-\tau
^{m+1}\right) -\nonumber\\
&&\tau _{0}\,_{2}F_{1}\left( \frac{1}{2},\frac{1}{m+1};1+\frac{1%
}{m+1};-\tau _{0}^{m+1}\right) \approx   
\left( \tau -\tau _{0}\right) \times \nonumber\\
&&\Bigg[ _{2}F_{1}\left( \frac{1}{2},\frac{1%
}{m+1};1+\frac{1}{m+1};-\tau _{0}^{m+1}\right) +
\frac{1}{2}\left( -\tau
_{0}^{m+1}\right) {}^{-\frac{1}{m+1}}\times \nonumber\\
&&B_{-\tau _{0}^{m+1}}\left( 1+\frac{1}{%
m+1},-\frac{1}{2}\right) \Bigg] ,
\end{eqnarray}%
where $B_{z}(a,b)$ gives the incomplete beta function $B_z(a,b)=\int_0^z{t^{a-1}(1-t)^{b-1}dt}$. Therefore from Eq.~(%
\ref{100}) we obtain
\begin{eqnarray}
\tau  &\approx &\tau _{0}+\nonumber\\
&&\frac{\left[\left( c_{0}/b\tau _{0}\right)
^{(m-1)/2}+\tau _{0}\,_{2}F_{1}\left( \frac{1}{2},\frac{1}{m+1};1+\frac{1}{%
m+1};-\tau _{0}^{m+1}\right)\right]e^{-\frac{m-1}{m+3}\frac{Vf}{D_{0}}\xi } }{_{2}F_{1}\left( \frac{1}{2},\frac{1}{m+1};1+%
\frac{1}{m+1};-\tau _{0}^{m+1}\right) +
\frac{1}{2}\left( -\tau
_{0}^{m+1}\right) ^{-\frac{1}{m+1}}B_{-\tau _{0}^{m+1}}\left( 1+\frac{1}{%
m+1},-\frac{1}{2}\right) }
- \nonumber\\
&&\frac{\left( c_{0}/b\tau _{0}\right) ^{(m-1)/2}+\tau _{0}%
\,_{2}F_{1}\left( \frac{1}{2},\frac{1}{m+1};1+\frac{1}{m+1};-\tau _{0}
^{m+1}\right) }{_{2}F_{1}\left( \frac{1}{2},\frac{1}{m+1};1+\frac{1}{m+1}%
;-\tau _{0}^{m+1}\right) +
\frac{1}{2}\left( -\tau _{0}^{m+1}\right) {}^{-%
\frac{1}{m+1}}B_{-\tau _{0}^{m+1}}\left( 1+\frac{1}{m+1},-\frac{1}{2}\right)
}.\nonumber\\
\end{eqnarray}%
With the use of the above expression for $\tau $ we obtain the number cell density in the first order approximation as
\begin{eqnarray}
c(\xi ) &\approx &b\Bigg\{ \tau _{0}-\Psi +\Phi
e^{-\frac{m-1}{m+3}\frac{Vf}{D_{0}}\xi }\Bigg\} e^{-\frac{2}{m+3}\frac{Vf}{D_{0}}\xi }\times
\Bigg\{ \left( \frac{c_{0}}{b\tau _{0}}\right) ^{(m-1)/2}+\nonumber\\
&&\tau
_{0}\,_{2}F_{1}\left( \frac{1}{2},\frac{1}{m+1};1+\frac{1}{m+1};-\tau
_{0}^{m+1}\right) \left[ 1-e^{-\frac{m-1}{m+3}\frac{Vf}{D_{0}}\xi }\right] \Bigg\}
^{2/(m-1)},  \nonumber\\
\end{eqnarray}%
where
\begin{equation}
\Psi =\frac{\left( c_{0}/b\tau _{0}\right) ^{(m-1)/2}+\text{$\tau _{0}$}%
\,_{2}F_{1}\left( \frac{1}{2},\frac{1}{m+1};1+\frac{1}{m+1};-\text{$\tau _{0}
$}^{m+1}\right) }{_{2}F_{1}\left( \frac{1}{2},\frac{1}{m+1};1+\frac{1}{m+1}%
;-\tau _{0}^{m+1}\right) +\frac{1}{2}\left( -\tau _{0}^{m+1}\right) {}^{-%
\frac{1}{m+1}}B_{-\tau _{0}^{m+1}}\left( 1+\frac{1}{m+1},-\frac{1}{2}\right)
},
\end{equation}%
and
\begin{equation}
\Phi =\frac{\left( c_{0}/b\tau _{0}\right) ^{(m-1)/2}+\tau
_{0}\,_{2}F_{1}\left( \frac{1}{2},\frac{1}{m+1};1+\frac{1}{m+1};-\tau
_{0}^{m+1}\right) }{_{2}F_{1}\left( \frac{1}{2},\frac{1}{m+1};1+\frac{1}{m+1}%
;-\tau _{0}^{m+1}\right) +\frac{1}{2}\left( -\tau _{0}^{m+1}\right) {}^{-%
\frac{1}{m+1}}B_{-\tau _{0}^{m+1}}\left( 1+\frac{1}{m+1},-\frac{1}{2}\right)
},
\end{equation}%
respectively.

\begin{figure}
   \centering
  \includegraphics[width=8cm]{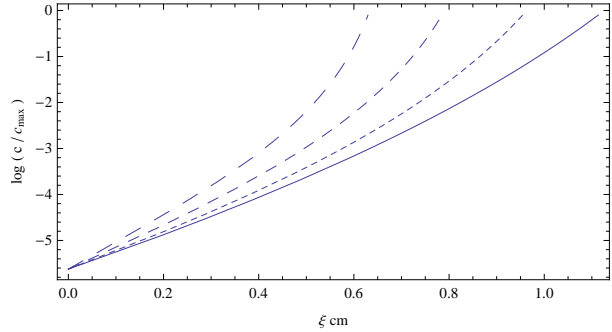}
  \caption{Variation of the ratio of the cell number density and the maximum cell number $c_{max}$, $c/c_{max}$, as a function of $\xi $ (in a logarithmic scale) for the second generalized Fisher-Kolmogorov model for $m =1.2$ and $A_m=10^{7.5}$ (solid curve), $m=1.3$ and $A_m=10^{6.94}$ (dotted curve), $m=1.4$ and $A_m=10^{6.4}$ (short dashed curve) and $m=1.5$ and $A_m=10^{5.87}$ (dashed curve), respectively. The diffusion coefficient $D_0=1.43$ cm$^2$/year, $V_f=-9$ cm/year, while the initial value of the parameter $\tau $ is $\tau _0=10$. The tumor concentration wave is travelling in the direction of positive $x$. }\label{fig3}
  \end{figure}

\begin{figure}
   \centering
  \includegraphics[width=8cm]{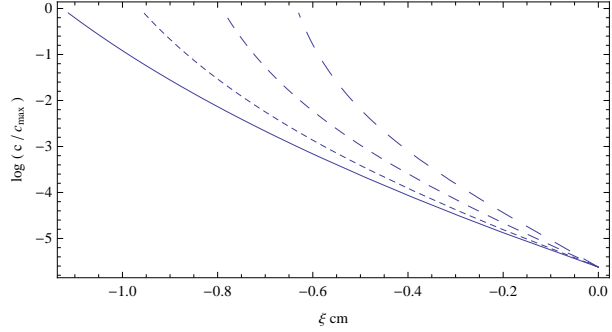}
  \caption{Variation of the ratio of the cell number density and the maximum cell number $c_{max}$, $c/c_{max}$, as a function of $\xi $ (in a logarithmic scale) for the second generalized Fisher-Kolmogorov model for $m =1.2$ and $A_m=10^{7.5}$ (solid curve), $m=1.3$ and $A_m=10^{6.94}$ (dotted curve), $m=1.4$ and $A_m=10^{6.4}$ (short dashed curve) and $m=1.5$ and $A_m=10^{5.87}$ (dashed curve), respectively. The diffusion coefficient $D_0=1.43$ cm$^2$/year, $V_f=9$ cm/year, while the initial value of the parameter $\tau $ is $\tau _0=10$. The tumor concentration wave is travelling in the direction of negative $x$. }\label{fig4}
  \end{figure}

The variation of the ratio of the cell number density divided by $c_{max}$ is represented, in a logarithmic scale, and for different values of $m$ and $A_m$ in Figs.~\ref{fig3} and \ref{fig4}, respectively.

\section{Discussions and final remarks}\label{sect5}

In the present paper we have investigated the travelling wave solutions of a general class of diffusion--reaction systems, with the diffusion and reaction functions given as functions of the concentration $c$ of the diffusing component. In this case the second order differential equation, describing the travelling wave solution for the system can be reduced to a first order first kind Abel ordinary non-linear differential equation. The integrability conditions of the Abel equation allow to obtain several classes of exact travelling wave solutions of the general reaction--diffusion system, with the reaction and diffusion functions satisfying some compatibility conditions. We have considered the solutions that can be found by using the Chiellini integrability condition, and the Lemke transformation, respectively. From the large class of exactly integrable reaction--diffusion equations we did concentrate on two equations, both representing generalizations of the Fisher--Kolmogorov equation with more general diffusion and reaction functions. More exactly, with the choices
\begin{equation}\label{120}
D(c)=\frac{D_0}{\left(1-c/c_{max}\right)^{\alpha}}, Q(c)=\rho c \left(1-\frac{c}{c_{max}}\right)^{\alpha},
\end{equation}
and
\begin{equation}
D(c)=D_0={\rm constant}, Q(c)=\frac{2m+1}{(m+3)^2}\frac{V_f^2}{D_0}c\left[1-A\left(\frac{c}{c_{max}}%
\right)^{m-1}\right],
\end{equation}
respectively, the travelling wave solutions of the corresponding one dimensional reaction-diffusion equations can be obtained in an exact parametric form. Both equations represent generalizations of the standard diffusion and Fisher--Kolmogorov equations, to which they reduce in some limiting case. For the first generalized Fisher--Kolmogorov equation, for $\alpha =0$ we reobtain the standard diffusion equation, while for $m=2$ the equation of the generalized tumor model growth reduces to the Fisher--Kolmogorov equation.

 From a very general physical point of view the behavior of microscopic particles is called ideal if they form very dilute solutions. In this limit, their positions, orientations, and movements are considered to be non-correlated \cite{bio}. The macroscopically measured quantities  usually are averages over an ensemble of molecules.  In real solutions, however, even weak inter-particle interactions will cause a concentration dependence
of the observed properties \cite{bio}. The study of such systems with weak inter-particle
interactions is important in many processes. For examples, weak interactions determine the tendency
for a suspension of particles to remain in solution, to aggregate, to overcome phase separation or to form crystals \cite{bio}. Concentration-dependent diffusion coefficients are also use to model mass transfer phenomena in polymer membranes \cite{bio1}, as well as water sorption in a polymer matrix composite \cite{bio2}. Several functional forms for the concentration dependence of the diffusion function have been proposed. A very simple such form is a linear dependence of the "long time" diffusion function $D(c)$, given by \cite{bio}
\begin{equation}
D(c)=D_0\left(1+k_Dc\right),
\end{equation}
where $D_0={\rm constant}$ is the "short time" diffusion coefficient, and $k_D$ is a constant that can be expressed quite generally in terms of the
pair correlation function of the particles, and the equilibrium segment distribution about their centre of mass.  A more general, both temperature $T$ and concentration dependent diffusion coefficient was considered in \cite{bio2}, with
\begin{equation}
D(T_b,T_a,c)=D_be^{E_a(c)/RT_b},
\end{equation}
where the activation energy $E_a(c)$ is given by
\begin{equation}
E_a(c)=\frac{R}{1/T_b-1/T_a}\ln\frac{\alpha +\beta c^{\gamma}}{D_b},
\end{equation}
with $R$ the ideal gas constant, $T_b$ and $T_a$ two distinct temperatures, and $D_b$, $\alpha $, $\beta $ and $\gamma $ constants.

We have also investigated the possibility of modelling of the glioblastoma tumor growth by using the obtained exact solutions of the general reaction-diffusion equation, corresponding to the specific choices of the diffusion and reaction functions. In order to obtain the numerical results we have used realistic values of the biological and physical parameters determining the tumor growth. Similarly the to Gompertz curve, for the first model the growth slows done at the end of a time period, with the density of the cells reaching a constant value. A very different behavior characterizes the second model, which for a monotonically increasing cell number density requires $V_f<0$, that is, $\xi =x+V_ft$. Moreover, for arbitrary $m$ the cell number density increases very rapidly with $\xi $. {\bf Therefore, this model is not relevant for the study of the glioblastoma growth, and its biological dynamic}. For $V_f>0$ we obtain a traveling wave solution of the second generalized Fisher--Kolmogorov equation traveling in the direction of negative $x$.

 In the present paper we have considered the biological implications of the travelling wave solutions of the generalized tumor growth equations, which corresponds to some given forms of the diffusion and reaction functions $D(c)$ and $Q(c)$. In the standard model of glioblastoma growth \cite{3}, two major biological phenomena underlying the growth of gliomas at the cellular scale are taken into account: proliferation and diffusion. The simplest choice for the proliferation term is a constant growth rate $\rho $, leading to an exponentially growing total number of glioma cells. For the invasive properties of gliomas, cell migration is assumed to be a random walk, corresponding to a passive (Fickian) diffusion characterized by a single coefficient $D$. Besides the logistic form $Q(c)=\rho c\left(1-c/c_{max}\right)$,  other forms for the reaction function $Q(c)$ have been  considered in the literature, like, for example, $Q(c)=\rho c\ln \left(c_m/c\right)$, corresponding to the Gompertz law \cite{Mar,MRI}. Improved mathematical models by including anisotropic extension of gliomas have been also investigated, by adopting a cell diffusion tensor derived from the water diffusion tensor, as given by MRI diffusion tensor imaging \cite{MRI}.

In our analysis of the tumor growth model we have extended the allowed functional forms  of $D(c)$ and $Q(c)$, and we have also considered the possibility of the existence of a functional dependence between the diffusion of the malignant cells and logistic growth.
In the first considered case, corresponding to the choices given by Eq.~(\ref{120}) for $D(c)$ and $Q(c)$, the product of these functions satisfy the relation $D(c)Q(c)=D_0\rho c$, that is, the product of the diffusion and reaction functions is proportional to the cell concentration. This implies a cell concentration dependence of the diffusion function of the form $D(c)\propto c/Q(c)$, which imposes a tight relation between $D(c)$ and $Q(c)$. From a biological point of view one can assume that at the beginning of the tumor evolution its growth is dominated by the logistic reaction function $Q(c)\neq 0$, so that $\lim_{c\rightarrow 0}D(c)=\lim_{c\rightarrow 0}c/Q(c)=0$, implying that at the early growth stages of the tumor one can neglect diffusion. On the other hand with the increase of the malign cell concentration, and with the decrease of $Q(c)$, the diffusion becomes the main physical process governing the spread of the tumor.

 The clinically distinct feature of glioblastoma lies within its infiltrative potential, rendering complete tumor resection nearly impossible \cite{bp1}. The glioblastoma cells have a  great migratory and invasive potential of their surroundings, which can be related mainly to their diffusive properties. In the varying diffusion and reaction cell growth model considered in the present papers this invasive potential can be enhanced, as compared to the standard Fisher--Kolmogorov model. As one can see from Fig.~\ref{fig1}, for $k=1/4$, the increase of the coefficient $\alpha $ in the diffusion and reaction functions leads to a significant increase in the size of the tumor, from around $\xi =0.3$ cm for $\alpha =1/2$ to $\xi =0.4$ cm for $\alpha =3/2$. Therefore the invasive properties of the glioblastoma cells are strongly dependent on the functional forms of $D$ and $Q$. On the other hand, in this case the maximum cell number is not affected by the variations of $\alpha $. A strong dependence of the invasiveness of glioblastoma cells on the numerical values of $k\neq 1/4$, which fixes the front velocity $V_f$,  can be observed in Fig.~\ref{fig2}. In this case, for fixed diffusion and reaction functions, the change in $k$ determines a large variation of both the tumor size and the maximum cell number. Therefore in this model the invasiveness of the glioblastoma cells with fixed $D$ and $Q$ is determined by the coefficient $k$. For the case of the growth models described by the second generalized Fisher--Kolmogorov equation, presented in Fig.~\ref{fig4}, since the diffusion coefficient is a constant, the tumor evolution and invasiveness is mostly determined by the reaction function $Q(c)$, whose dependence on the concentration is described by the parameter $m$, and the constant $A_m$. There is a very significant difference in invasiveness for the different models, with combinations of the free parameters resulting in   tumoral extensions as large as $\xi =1$ cm.

 In the present paper we have ignored the effects of the radiotherapy and chemotherapy on the glioblastoma growth and evolution, described by Eqs.~(\ref{chem}) and (\ref{rad}), respectively. In these cases the presence of the explicitly time and space dependent terms $G(t)$ and $R\left(\vec{r},t\right)$ do not allow, in general, the reduction of the corresponding reaction--diffusion equations to an Abel type equation, unless some very particular functional forms of $G$ and $R$ are assumed.

Most of the mathematical studies of the reaction-diffusion equations, including those modelling the growth of glioblastomas, have been done by using either qualitative methods, or asymptotic evaluations. The usual way to study nonlinear
reaction--diffusion equations similar to Eq.~(\ref{3}) consists of the analysis of the behavior of the system on a phase plane
$(dc/d\xi, c)$, useful for qualitative analysis, however insufficient for finding any exact solution or testing a numerical
one. Finding exact analytical solutions of the tumor growth models can considerably simplify the process of comparison of the theoretical predictions with the experimental data, without the need of using complicated numerical procedures.

\section*{Acknowledgments}

We would like to thank to the two anonymous reviewers for comments and suggestions that helped us to significantly improve our manuscript.

\medskip
Received xxxx 20xx; revised xxxx 20xx.
\medskip

\end{document}